\documentstyle[12pt,epsf]{article}
\begin{document}
\def\gsim{{~\raise.15em\hbox{$>$}\kern-.85em \lower.35em\hbox{$\sim$}~}}
\def\lsim{{~\raise.15em\hbox{$<$}\kern-.85em \lower.35em\hbox{$\sim$}~}}
\def\C{p_{\mu\times\mu}}
\begin{titlepage}
\begin{flushright}
        BINP-97-11\\
        ISU-HET-97-5 \\
        October 1997
\end{flushright}
\vfill
\begin{center}
{\large \bf $CP$ Violating Anomalous Couplings 
at a $500$~GeV $e^+e^-$ Linear Collider}
\vfill
	{\bf A.~A.~Likhoded$^{(a)}$},
        {\bf G.~Valencia$^{(b)}$} 
        {\bf  and O.P.~Yushchenko$^{(c)}$}\\
{\it  $^{(a)}$ Branch of The Institute for  Nuclear Physics,\\ 
Protvino,  142284 Russia}. E-mail: likhoded@mx.ihep.su \\
{\it  $^{(b)}$ Department of Physics, Iowa State University, Ames IA 50011}\\
E-mail: valencia@iastate.edu \\
{\it  $^{(c)}$ Institute for High Energy Physics,\\ 
Protvino,  142284 Russia}. E-mail: yushchenko@mx.ihep.su \\
\vfill
\end{center}
\begin{abstract}

We study the sensitivity of a $500$~GeV $e^+e^-$ Linear Collider to 
$CP$ violating anomalous couplings. We find that with 50~fb$^{-1}$, 
and taking only one non-zero coupling at a time,  
the process $e^+e^-\rightarrow W^+ W^-$ can be 
used to place the 95\% confidence level bounds 
$|\tilde\kappa_\gamma|\leq0.1$, $|\tilde\kappa_Z|\leq 0.1$ and 
$|g_4^Z|\leq 0.1$ from $CP$ even observables. By studying  
certain distributions in the process 
$e^+e^-\rightarrow \mu^+\mu^-\nu\overline{\nu}$ one of the bounds can  
be improved to $|g_4^Z| \leq 0.06$. This process also allows 
the construction of a $CP$ odd observable which 
can be used to place bounds on $CP$ violating new physics. At 
the 95\% confidence level we find 
$|\tilde\kappa_\gamma|\leq 0.3$,  $|\tilde\kappa_Z|\leq 0.2$ 
and a much weaker bound for $|g_4^Z|$.

\end{abstract}

\end{titlepage}

\section{Introduction}

In spite of its remarkable phenomenological success there 
are several aspects of the standard model that remain unexplained. 
Two of them 
are the mechanism of electroweak symmetry breaking and the origin of 
$CP$ violation. A broad class of models in which the electroweak symmetry 
is broken dynamically by new strong interactions does not contain 
any new particles sufficiently light to be produced at a 500~GeV 
$e^+e^-$ collider. New particles in these models have masses in the TeV range 
and only manifest themselves indirectly in experiments at lower energies. 
In general these models may violate $CP$ and this would also manifest 
itself indirectly at low energy.

It is convenient to describe the phenomenology of the most important 
features of this type of new physics in a model independent way. 
This is accomplished 
by studying a low energy (below a few TeV) effective Lagrangian that 
contains only the standard model fields and where the effect of the 
new physics appears as higher dimension operators. These higher dimension 
operators modify the couplings of the observed particles, inducing 
``anomalous couplings'' whose phenomenology has been 
studied in detail \cite{review}. 

In this paper we study the effect of the lowest dimension 
operators that violate $CP$ in the gauge-boson self-couplings. 
In has become standard to parameterize the three gauge boson 
coupling $WWV$ following the notation of Ref.~\cite{hagi}:
\begin{equation}
{\cal L}_{WWV} = g_{WWV}\biggl(-g_4^V W^+_\mu W^-_\nu
(\partial^\mu V^\nu +\partial^\nu V^\mu ) + {i\over 2} \tilde{\kappa}_V
\epsilon_{\mu \nu \alpha \beta}W^+_\mu W^-_\nu V^{\alpha \beta}\biggr)
\label{haginot}
\end{equation} 
In writing this equation we have already dropped terms proportional to 
$\tilde{\lambda}_V$ because they are of higher dimension. The 
overall normalization is 
$g_{WW\gamma} =  -e$ and $g_{WWZ} = -e\cot\theta _W$. Electro-magnetic 
gauge invariance forbids the term $g_4^\gamma$ so we are left with 
three new $CP$ violating parameters. One of them, $g_4^Z$, violates $C$  
and conserves $P$; whereas the other two, 
$\tilde{\kappa}_\gamma$ and $\tilde{\kappa}_Z$ violate $P$ and conserve $C$. 

The next to leading order electroweak chiral Lagrangian \cite{ewcl} 
contains three $CP$ violating operators whose couplings correspond to 
the lowest dimension contributions to the parameters 
in Eq.~(\ref{haginot}). They are \cite{appel}: 

\begin{eqnarray}
{\cal L}_{CP-odd} &=&  2 \alpha_{12} g 
{\rm Tr}(T{\cal V}_\mu){\rm Tr}({\cal V}_\nu W^{\mu\nu}) \nonumber \\
&+& {1 \over 4}\alpha_{13} g g^\prime 
\epsilon^{\mu\nu\rho\sigma}B_{\mu\nu}{\rm Tr}(TW_{\rho\sigma})
\nonumber \\
&+& {1 \over 8}\alpha_{14}g^2
\epsilon^{\mu\nu\rho\sigma}{\rm Tr}(TW_{\mu\nu}){\rm Tr}(TW_{\rho\sigma})
\label{chilag}
\end{eqnarray}
where we have used the notation of Ref.~\cite{appel}.
The relation between these couplings and those in Eq.~(\ref{haginot}) 
is $\tilde{\kappa}_Z = e^2(\alpha_{13}/c^2_W
-\alpha_{14}/s^2_W)$, $\tilde{\kappa}_\gamma =- 
(e^2/s^2_W)(\alpha_{13}+\alpha_{14})$ and $g_4^Z = -
e^2/(s^2_Wc^2_W)\alpha_{12}$ \cite{appel}.

We do not wish to reproduce here all the details of the notation of 
Ref.~\cite{appel}, it is sufficient to remind the reader that the factor 
$T$ that appears in all three operators constitutes an explicit 
breaking of custodial symmetry. This implies that if the new physics 
violates $CP$ maximally (i.e. without suppressions from dimensionless 
parameters such as mixing angles), the natural size of the coefficients 
is that of $(v^2/\Lambda^2) \Delta\rho$. These are the same arguments 
that give the natural size of the lowest dimension parity violating 
coupling \cite{parvio}. With $v \approx 246$~GeV and the scale of new physics 
$\Lambda$ a few TeV we thus expect $\alpha_{12,13,14}\sim 10^{-4}$ if 
the symmetry breaking sector has a custodial symmetry to explain the 
smallness of $\Delta\rho$. In Ref.~\cite{appel} 
Appelquist and Wu discuss a specific model in which they estimate that 
the coefficients $\alpha_{12,13,14}$ are indeed of order $10^{-4}$ 
and correlated with $\Delta\rho$. On the other hand, if $\Delta\rho$ 
is small accidentally, naive power counting tells us that these 
couplings could be at the few percent level, ${\cal O}(v^2/\Lambda^2)$. 
These numbers will help us calibrate the significance 
of the constraints that we discuss.

The best indirect bound that exists on any of these couplings is 
$|\tilde{\kappa}_\gamma|< 2 \times 10^{-4}$, which arises from the 
neutron e.d.m. \cite{neutron}. This is a very tight constraint, but it 
is subject to naturalness assumptions. As usual, it is not a substitute 
for a direct constraint. Previous studies of $W\gamma$ production at 
an upgraded Tevatron have concluded that it will be possible to 
place the constraint $|\tilde{\kappa}_\gamma| \lsim 0.1$ \cite{tevabound}.
This is of the same order as the bound that we find in this study for the 
500~GeV $e^+e^-$ collider, but a precise comparison 
is not possible without further knowledge of the experimental setups. 

There have been previous studies of the $CP$ violating anomalous couplings 
in the process $e^+e^- \rightarrow W^+ W^-$ \cite{previous}, 
but a detailed numerical analysis of the bounds that one can get at a
NLC has not been done. The process
$e^+e^- \rightarrow \nu \overline{\nu}Z$ has also been recently considered 
\cite{rindani}. The authors of Ref.~\cite{rindani} find that one could 
place the bound $g_4^Z \lsim 0.1$ by studying a forward-backward asymmetry 
with $50~{\rm fb}^{-1}$.

\section{Bounds from observables in $e^+e^- \rightarrow W^+ W^-$}

We start with the process 
$e^+ e^- \rightarrow W^+ W^-$ at a center of mass energy of 500~GeV 
without considering any specific decay channel for the $W$~bosons.  
At this stage we also ignore the $CP$ violating nature of the couplings 
$\tilde\kappa_\gamma,~\tilde\kappa_Z,~g_4^Z$ and look only at the 
quadratic effects that they induce in the decay distribution. It is 
possible to study truly $CP$-violating effects that are linear in the 
couplings in one of two ways. We could include absorptive phases in 
the form factors associated with Eq.~(\ref{haginot}) \cite{hagi,appel}. 
If these phases arise from the same sector responsible for the 
anomalous couplings then they do not introduce additional suppression 
factors. This can be seen, for example, in the model of Ref.~\cite{appel}.
Alternatively, we could construct a $CP$ odd observable involving the 
polarization vectors of the $W$ bosons. This is equivalent to studying 
correlations that involve the momenta of the decay products of the 
$W$-bosons to a specific channel. We take the second approach in the 
following section.

In this section we consider the $W$~bosons to be final state particles and 
take into account the efficiency for $W^+W^-$ pair reconstruction 
$\epsilon_{WW}=0.15$ \cite{effi} in our numerical simulation. 
Because we are ignoring, for now, the $CP$ violating nature of the 
couplings, it is possible to bound them using the 
same $CP$ even observables that we studied in Ref.~\cite{lihan}. 
The only difference between the $CP$ violating couplings that we
study here, and the $CP$ conserving couplings that we studied in 
Ref.~\cite{lihan} is that the $CP$ violating couplings always appear 
quadratically in the $CP$ even distributions used to place the 
bounds.\footnote{Again, this is because we are not including 
any possible absorptive phases. We can justify this {\it a posteriori} 
because the bounds that can be obtained from terms linear in the 
couplings and terms quadratic in the couplings are very similar 
due to the low statistics.} 
We use the same assumptions about systematic uncertainties  
and the same analysis of the differential distribution 
that we described in detail in Ref.~\cite{lihan}. 
In particular, we use a systematic error $\sim$1.5\%. This  
number arises from 
an uncertainty in the luminosity measurement of 
$\simeq$0.5\%, an error in the acceptance $\simeq$1\%, 
an error for background subtraction $\simeq$0.5\% 
and a systematic error in the knowledge of the branching ratio
$\simeq$0.5\%. From  
a $\chi^2$ analysis of $d\sigma/d\cos\theta$ with 5 bins, 
we find the 95\% confidence level bounds 
(taking only one non-zero coupling at a time): 
\begin{equation}
|\tilde\kappa_\gamma| \leq  0.1 ,\;
|\tilde\kappa_Z| \leq  0.1 , \;
|g_4^Z| \leq  0.09
\label{boundsfromww}
\end{equation}
The best bounds for this process are obtained using four bins, as 
discussed in Ref.~\cite{lihan}. However, the bounds we obtain using 
five bins are indistinguishable from those for four bins. We prefer 
five bins because this will be the optimal number for 
$e^+e^- \rightarrow \mu^+ \mu^- \nu \overline{\nu}$ and using the 
same number of bins in both cases will facilitate a comparison. 

\begin{figure}[htb]
\centerline{\epsfxsize=2.0in\epsfbox{fig1a.eps}\hspace{0.2in}
\epsfxsize=2.0in\epsfbox{fig1b.eps}}
\centerline{\epsfxsize=2.0in\epsfbox{fig1c.eps}}
\caption[figure 1]{Allowed (95\% C. L.) region  from 
the $CP$ even angular distribution of $e^+e^- \rightarrow W^+ W^-$ for: 
a) $\tilde{\kappa}_Z~-~g_4^Z$ with $\tilde{\kappa}_\gamma=0$; 
b) $\tilde{\kappa}_\gamma~-~g_4^Z$ with $\tilde{\kappa}_Z=0$; 
c) $\tilde{\kappa}_\gamma~-~\tilde{\kappa}_Z$ with $g_4^Z=0$.}
\label{f:wwtwo}
\end{figure}
In Figure~\ref{f:wwtwo} we 
present the allowed 95\% confidence level regions when we take 
one of the $CP$ violating couplings to be zero. 
We see that the bounds are indeed similar to those  that can be 
placed at an upgraded Tevatron \cite{tevabound}, and of the same order
as the bounds that we found for $CP$ conserving anomalous 
couplings \cite{lihan}. The fact that we obtain similar bounds 
for couplings that contribute linearly to the differential 
cross-section and for couplings that only contribute quadratically, 
already indicates that this process is not sensitive to the very 
small values predicted by naive dimensional analysis.

\section{Bounds from observables in $e^+e^- \rightarrow \mu^+ \mu^- 
\nu \overline{\nu}$}

We now wish to consider a specific channel for the decay of 
the $W$-bosons so that we can construct correlations that 
could single out $CP$-violating interactions. With this in 
mind, we need a final state that is easy to identify and that 
transforms into itself under $CP$. 
We thus choose to identify the $W$ pairs by their $\mu \nu$ leptonic 
decays. We calculate the amplitudes for the process 
$e^+e^-\to \mu^+\mu^-\nu\bar\nu$ and generate events for the 
three following subprocesses:
\begin{eqnarray}
\mbox{Subprocess I } &\mbox{---}& 
e^+e^-\to \mu^+\mu^-\nu_{\mu}\bar\nu_{\mu}\:, \nonumber\\
\mbox{Subprocess II } &\mbox{---}& 
e^+e^-\to \mu^+\mu^-\nu_{e}\bar\nu_{e}\;, \nonumber\\
\mbox{Subprocess III } &\mbox{---}& 
e^+e^-\to \mu^+\mu^-\nu_{\tau}\bar\nu_{\tau}\;. \nonumber
\end{eqnarray}
These subprocesses include 20, 21 and 11 Feynman diagrams respectively. 
The last one does not contain anomalous vertices, and constitutes  
pure background. Sufficient events are generated with our Monte-Carlo 
to achieve a $1\%$ statistical error in the value of the cross-section. 

We first generate events with a cut on the muon scattering 
angle and on the muon pair invariant mass:
\begin{equation}
170^o \leq \theta\leq 10^o,~~~  M_{\mu\mu} \geq 30~{\rm GeV}
\label{firstcuts}
\end{equation}
The angle $\theta$ is the scattering angle between the $\mu^-$ 
and the $e^-$ momenta in the $e^+e^-$ center of mass frame. To study 
$CP$ odd observables we need to make sure that our cuts are 
``$CP$ blind'', so the same cut is imposed on the angle 
between the $\mu^+$ and the $e^+$ momenta. 
These cuts are similar to the ones used by the
LEP experiments.\footnote{The current experiments at LEP have 
a typical region for muon reconstruction of $170^o < \theta < 10^o$. 
We assume that the experiments at a NLC will
have a similar geometry so we use the same cut. The cut on the invariant
mass of $\mu^+ \mu^-$ pair,  $m_{\mu\mu} > 30$ GeV serves to reject 
Dalitz conversion of soft photons and to insure good angular 
separation of the muons.} 
After imposing these  cuts we assume a muon reconstruction efficiency 
equal to 1. 

The total cross-section for the $e^+e^-\to \mu^+\mu^-\nu\bar\nu$ process 
with these cuts is 7.65 fb, 
which results (with an integrated luminosity of 50 fb$^{-1}$) 
in only 382 events. The resulting bounds will be limited by the small 
statistics so we will have to relax these cuts later on. 

We start by placing bounds on the $CP$ violating anomalous couplings 
using the following observables:
\begin{equation}
\frac{d\sigma}{d\cos\theta}\:,\;\; \frac{d\sigma}{dp_{\mu}}\:,\;\;
\frac{d\sigma}{d\C}\:, 
\end{equation}
where $\theta$ is again the scattering angle between the $\mu^-$ 
and the $e^-$ momenta in the $e^+ e^-$ center of mass frame; 
$p_{\mu}$ is the muon three-momentum in the same frame, and $\C$ is 
proportional to the $T$-odd correlation 
$\vec p_e \cdot [\vec p_{\mu^+}\times \vec p_{\mu^-}]$. Numerically 
we work in the $e^+e^-$ center of mass frame and use: 
\begin{equation}
\C = \frac{4}{s}[\vec p_{\mu^+}\times \vec p_{\mu^-}]_z \;,
\label{defofc}
\end{equation}
where the index $z$ denotes the component along the beam direction. 
With this normalization $\C$ can take values from $-1$ to 1. 
If the polarization of the final leptons is not observed, and the 
beams are not polarized, this correlation serves to analyze the 
$CP$ properties of the interaction \cite{donval}. $CP$ even 
interactions give rise to symmetric distributions (symmetric about the
point $\C=0$) in $\C$, whereas $CP$ odd interactions give rise 
to antisymmetric  distributions in $\C$. The antisymmetric distributions in  
this correlation will arise from interference between standard model 
amplitudes and the new $CP$-violating physics and will be linear in the new 
couplings. Notice that the correlation $\C$ is also odd under parity. 
This means that we will get terms proportional to
$\tilde{\kappa}_{\gamma,Z}$ from interference with the parity even 
standard model amplitude and a term proportional to $g_4^Z$ from 
interference with the parity odd standard model amplitude. 

In Figure~\ref{f:smdist} we show the differential cross-section predicted 
by the standard model at lowest order, as a function 
of $\cos\theta$ and $p_{\mu}$,  for $\sqrt{s}=500$~GeV. 
From these figures we see that the events predicted by the standard model 
are concentrated at small scattering angles and low muon momentum. 
Similarly we find that the standard model events have a symmetric 
distribution in $\C$ (as corresponds to CP conservation) that is very 
strongly peaked at $\C=0$. 

\begin{figure}[htb]
\centerline{\epsfxsize=2.0in\epsfbox{fig2a.eps}\hspace{0.2in}
\epsfxsize=2.0in\epsfbox{fig2b.eps}}
\caption[figure 2]{Standard model differential cross-section for the 
process $e^+e^- \rightarrow \mu^+ \mu^- \nu \overline{\nu}$ at 
$\sqrt{s}=500$~GeV with the cuts of Eq.~(\ref{firstcuts}) 
as a function of a) $\cos\theta$ and b) $p_\mu$. }
\label{f:smdist}
\end{figure}

In order to understand the effect of the cuts that we impose on 
the distribution with respect to $\C$ it is convenient to 
express this correlation analytically. In addition to the angle 
$\theta$ defined above, we need to define 
$\theta_\mu$, the angle between the $\mu^+$ and the $\mu^-$ 
momenta in the $e^+e^-$ center of mass frame. Using a 
coordinate system with the $z$ axis pointing in the 
direction of the $\mu^-$ momentum, the $e^-$ momentum being in the 
$x-z$ plane, and with $\phi_\mu$ being  the azimuthal angle of the 
$\mu^+$ in this coordinate system, the correlation is proportional 
to:
\begin{equation}
\C \sim \sin\theta \sin\theta_\mu \sin\phi_\mu
\label{anformofc}
\end{equation}
Working in the limit of massless muons, the angle $\theta_\mu$ 
is related to the invariant mass of the muon pair in the 
following way:
\begin{equation}
\cos\theta_\mu = 1 - {M^2_{\mu\mu}\over 2E_{\mu^+}E_{\mu^-}}
\label{angleinvm}
\end{equation}
We can now understand the effect of the cuts of Eq.~(\ref{firstcuts}): 
a) they remove the region of small $\sin\theta$ where $\C$ is small, 
thus increasing the signal to background ratio; b) they remove the 
region of small muon pair invariant mass effectively enhancing 
the region where $\cos\theta_\mu$ is negative. 

In Figure~\ref{f:chdist} we show the deviations 
induced by the $CP$-violating couplings in the 
differential cross-section. For illustration purposes  
we use the values  $g_4^Z=0.1$, 
$\tilde\kappa_{Z}=0.5$, and 
$\tilde\kappa_{\gamma}=0.5$, with only one of them 
being non-zero at a time. The curves labeled 
`1',  `2', and `3' correspond to the case of non-zero  
$g_4^Z$, $\tilde\kappa_{Z}$, and $\tilde\kappa_{\gamma}$ respectively. 
It is clear from these figures that the kinematic regions where the 
new effects would be most important are high muon momentum and backward 
scattering. The distribution with respect to $\C$, is approximately 
symmetric about $\C=0$ indicating that for the cuts in
Eq.~(\ref{firstcuts}), the terms quadratic in the new couplings (and thus 
$CP$-even) dominate.

\begin{figure}[htb]
\centerline{\epsfxsize=2.0in\epsfbox{fig3a.eps}\hspace{0.2in}
\epsfxsize=2.0in\epsfbox{fig3b.eps}}
\centerline{\epsfxsize=2.0in\epsfbox{fig3c.eps}}
\caption[figure 3]{Deviations in differential cross-section  
with the cuts of Eq.~(\ref{firstcuts}) from their 
standard model values as a function of a) $\cos\theta$, b) $p_\mu$ 
and c) $\C$. In all cases 
the curves labeled 1, 2 and 3 correspond to $g_4^Z=0.1$, 
$\tilde{\kappa}_Z = 0.5$ and $\tilde{\kappa}_\gamma=0.5$ respectively.}
\label{f:chdist}
\end{figure}

\begin{figure}[htb]
\centerline{\epsfxsize=2.in\epsfbox{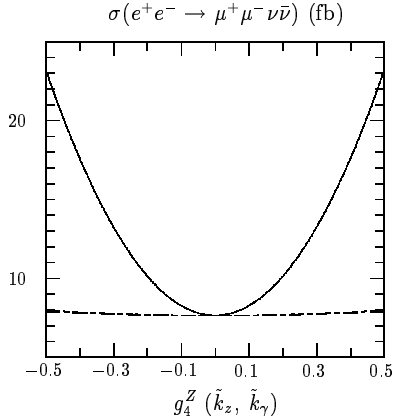}}
\caption[figure 4]{Total cross-section with the cuts 
of Eq.~(\ref{firstcuts}) as a function of $g_4^Z$ (solid line) and
$\tilde{\kappa}_{Z,\gamma}$ (dashed-lines indistinguishable in this figure).}
\label{f:totsig}
\end{figure}

It is interesting to notice that for the set of cuts that we have used
so far, 
the total cross-section is more sensitive to the value of $g_4^Z$ than 
any of the distributions, we show this in Figure~\ref{f:totsig}. 

To place bounds on the anomalous couplings, we used a standard 
$\chi^2$-criterion to analyze the events (including the 0.5\% 
anticipated systematic error in the luminosity measurement). 
We also investigated the sensitivity of the resulting bounds to 
different kinematic cuts and binning, but we did not find any  
way to enhance the sensitivity to the new couplings. This is probably 
due to the very low statistics available (382 events). We find that 
the best bounds are obtained by dividing the events into 5 bins. 

The results for $\sqrt{s}=500$~GeV, integrated luminosity of 
50 fb$^{-1}$, 5 bins, and at the 95\% C.L. are shown in 
Figure~\ref{f:contfirstcut}.  
In this figure the solid contours correspond to the bounds coming  
from the muon momentum distribution, the short-dashed contours  
from the scattering angle distribution,
and the long-dashed contours from the correlation $\C$.  
For $\tilde{\kappa}_{\gamma,Z}$ the best bounds arise from the muon 
momentum distribution, 
at about the same level as the bounds that this process 
places on $CP$ conserving anomalous couplings \cite{likh}. For $g_4^Z$ 
the bound is $g_4^Z \leq 0.06$, slightly better than what we got in 
Eq.~(\ref{boundsfromww}), whereas for 
$\tilde{\kappa}_{Z,\gamma}$ the bounds are worse than those in 
Eq.~(\ref{boundsfromww}).
\begin{figure}[htb]
\centerline{\epsfxsize=2.0in\epsfbox{fig5a.eps}\hspace{0.2in}
\epsfxsize=2.0in\epsfbox{fig5b.eps}}
\centerline{\epsfxsize=2.0in\epsfbox{fig5c.eps}}
\caption[figure 5]{95\% C.L. bounds taking one coupling to be zero at 
a time and using the cuts of Eq.~(\ref{firstcuts}). 
The solid contours correspond to the bounds coming  
from the $p_\mu$ distribution, the short-dashed contours  
from the $\cos\theta$ distribution, 
and the long-dashed contours from the $\C$ distribution.} 
\label{f:contfirstcut}
\end{figure} 

Since our bounds are probably limited by the low statistics, we now 
study the effect of relaxing the cuts, and impose 
only the minimal cut:
\begin{equation}
M_{\mu^+\mu^-}\geq 5\;~{\rm GeV}.
\label{mincut}
\end{equation}
This is a very optimistic cut that still permits high efficiency in 
muon detection. Relaxing the cuts in this way has the effect 
of increasing the cross-section 
for the process $e^+e^-\to \mu^+\mu^-\nu\bar\nu$ 
to about 113 fb. With an integrated luminosity of 50 fb$^{-1}$ this 
results in 5660 events and consequently much better statistics.  

\begin{figure}[htb]
\centerline{\epsfxsize=2.in\epsfbox{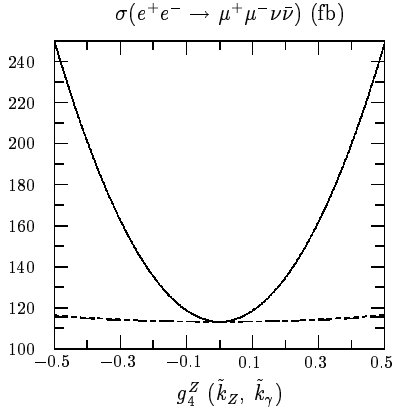}}
\caption[figure 6]{Total cross-section with the cuts 
of Eq.~(\ref{mincut}) as a function of $g_4^Z$ (solid line) and
$\tilde{\kappa}_{Z,\gamma}$ (dashed-lines indistinguishable in this figure).}
\label{f:sigtotn}
\end{figure}

In Figure~\ref{f:sigtotn} we present the total cross-section as a
function of the anomalous coupling $g_4^Z$ (solid line),  
$\tilde\kappa_{Z}$ (short-dashed line) and 
$\tilde\kappa_{\gamma}$ (long-dashed line). Comparison with 
Figure~\ref{f:totsig} shows that the relaxed cuts increase the 
sensitivity to the anomalous couplings.
In Figure~\ref{f:smdiffn} we show the differential 
distributions with respect to the 
muon scattering angle, $\cos\theta$, and the muon momentum, $p_{\mu}$. 
\begin{figure}[htb]
\centerline{\epsfxsize=2.0in\epsfbox{fig7a.eps}\hspace{0.2in}
\epsfxsize=2.0in\epsfbox{fig7b.eps}}
\caption[figure 7]{Standard model differential cross-section for the 
process $e^+e^- \rightarrow \mu^+ \mu^- \nu \overline{\nu}$ at 
$\sqrt{s}=500$~GeV and the cuts in Eq.~(\ref{mincut}) 
as a function of a) $\cos\theta$ and b) $p_\mu$.}
\label{f:smdiffn}
\end{figure} 
Once again we see that the standard model populates the regions of 
small scattering angle and low muon momentum preferentially.
In Figure~\ref{f:chdistn} we show the change in the differential 
cross-section when the anomalous couplings take values 
$g_4^Z=0.1$, $\tilde\kappa_{Z}=0.5$, and $\tilde\kappa_{\gamma}=0.5$. 
We take only one non-zero anomalous coupling at a time and use 
these values for illustration purposes only.
\begin{figure}[htb]
\centerline{\epsfxsize=2.0in\epsfbox{fig8a.eps}\hspace{0.2in}
\epsfxsize=2.0in\epsfbox{fig8b.eps}}
\centerline{\epsfxsize=2.0in\epsfbox{fig8c.eps}}
\caption[figure 8]{Deviations in differential cross-section 
with the cuts of Eq.~(\ref{mincut}) from their 
standard model values as a function of a) $\cos\theta$, b) $p_\mu$ 
and c) the $CP$ odd correlation of Eq.~(\ref{defofc}). In all cases 
the curves labeled 1, 2 and 3 correspond to $g_4^Z=0.1$, 
$\tilde{\kappa}_Z = 0.5$ and $\tilde{\kappa}_\gamma=0.5$ respectively.}
\label{f:chdistn}
\end{figure}
The standard model distribution is largest near $\theta=0$ whereas 
the new physics contributions are largest near $\theta=90^o$. 
Nevertheless, we find better sensitivity to the new physics when we 
do not impose the angular cut that excludes the region of small 
$\theta$, indicating that our analysis is limited by statistics.  

To place bounds on the anomalous couplings, we 
use a standard $\chi^2$-criterion to analyze the events, include the
anticipated 0.5\% systematic error in the luminosity measurement, and  
take the muon identification efficiency to be 1. We find that the 
best bounds are achieved by dividing the events into 5 bins.  
The 95\% C.L. results for $\sqrt{s}=500$~GeV, integrated luminosity 
of 50 fb$^{-1}$ and 5 bins for the bounds on 
$g_4^Z$, $\tilde\kappa_Z$, and  $\tilde\kappa_{\gamma}$ 
are shown in Figure~\ref{f:contmincut}. In this figure we have taken one 
of the three couplings to be zero and looked at the projection of the 
allowed region into the plane of the other two couplings.
\begin{figure}[htb]
\centerline{\epsfxsize=2.0in\epsfbox{fig9a.eps}\hspace{0.2in}
\epsfxsize=2.0in\epsfbox{fig9b.eps}}
\centerline{\epsfxsize=2.0in\epsfbox{fig9c.eps}}
\caption[figure 9]{95\% C.L. bounds taking one coupling to be zero at 
a time and using the cuts of Eq.~(\ref{mincut}). 
The short-dashed contours correspond to the bounds coming  
from the $p_\mu$ distribution, the long-dashed contours  
from the $\cos\theta$ distribution, 
and the solid contours the $\C$ distribution.} 
\label{f:contmincut}
\end{figure} 
In this figure the solid contours correspond to the bounds 
coming from the correlation $\C$, 
the short-dashed contours  from the muon momentum 
distribution, and the long-dashed contours from the scattering 
angle distribution. 
Comparing Figures~\ref{f:contfirstcut}~and~\ref{f:contmincut} 
one can see that with the 
relaxed cuts, the best bounds are obtained from the correlation 
$\C$. The improvement in the bounds is partly due to the increased 
statistics, and mostly due to the fact that when we relax the cut on 
the muon pair invariant mass we include a region of phase space 
that has the largest sensitivity to $\C$. In view of 
Eqs.~(\ref{anformofc})~and~(\ref{angleinvm}), the region of smaller muon pair 
invariant mass ($M_{\mu^+\mu^-}<30$~GeV) appears to contain the 
region where $\sin\theta_\mu$ is large.

Alternatively, taking only one non-zero coupling at a time we find 
(from the correlation of Eq.~(\ref{defofc})):
\begin{equation}
|\tilde\kappa_\gamma| \leq  0.27 ,\;
|\tilde\kappa_Z| \leq  0.18 ,\;
|g_4^Z| \leq  0.08.
\end{equation}

\section{Bounds from a $CP$-odd observable}

In the previous section we have seen that the bounds that can be 
placed on the anomalous couplings using the correlation $\C$ are 
of the same order as those that can be placed from other observables. 
Therefore, it is interesting to see whether one can isolate the 
$CP$-odd components of the distributions with respect to $\C$, 
and in that way be able to really bound new $CP$ violating interactions.

To understand the effect of the different sets of cuts on these 
bounds we present in Figures~\ref{f:cpoddg}~and~\ref{f:cpoddz} the 
differential cross-section with respect to the correlation $\C$ 
for $\tilde{\kappa}_\gamma=0.5$ and $\tilde{\kappa}_Z=0.5$ 
respectively. 

\begin{figure}[htb]
\centerline{\epsfxsize=2.0in\epsfbox{fig10a.eps}\hspace{0.2in}
\epsfxsize=2.0in\epsfbox{fig10b.eps}}
\caption[figure 10]{Deviations in the differential cross-section with 
a) the cuts of Eq.~(\ref{firstcuts}) and b) the cuts of Eq.~(\ref{mincut}) 
as a function of $\C$ for $\tilde{\kappa}_\gamma = 0.5$. We have
separated the contribution from the term linear in
$\tilde{\kappa}_\gamma$ (antisymmetric curve) from that due to the 
term quadratic in $\tilde{\kappa}_\gamma$ (symmetric curve).}
\label{f:cpoddg}
\end{figure} 

\begin{figure}[htb]
\centerline{\epsfxsize=2.0in\epsfbox{fig11a.eps}\hspace{0.2in}
\epsfxsize=2.0in\epsfbox{fig11b.eps}}
\caption[figure 11]{Deviations in the differential cross-section with 
a) the cuts of Eq.~(\ref{firstcuts}) and b) the cuts of Eq.~(\ref{mincut}) 
as a function of $\C$ for $\tilde{\kappa}_Z = 0.5$. We have
separated the contribution from the term linear in
$\tilde{\kappa}_Z$ (antisymmetric curve) from that due to the 
term quadratic in $\tilde{\kappa}_Z$ (symmetric curve).}
\label{f:cpoddz}
\end{figure} 

In these 
figures we have separated the contributions to the differential 
cross-section arising from terms linear in the anomalous couplings 
(the curves that are antisymmetric about $\C=0$) from those arising from 
terms quadratic in the anomalous couplings (the curves that are symmetric 
about $\C=0$). We also show how these results vary when we go from 
the stronger cuts of Eq.~(\ref{firstcuts}) to the relaxed cut of 
Eq.~(\ref{mincut}).

Notice that the normalization of the curves with strong and weak 
cuts is different as it corresponds to the respective total
cross-section. 
From these curves we see that the relaxed cuts are not only better 
because they increase the statistics, but they also increase the 
relative contribution of the truly $CP$-odd term linear in 
$\tilde{\kappa}_{\gamma,Z}$ as we argued in the previous section. 

In order to quantify the bounds that can be placed in this way, 
we introduce the integrated $CP$-odd observable 
\begin{equation}
A\equiv \int {d\sigma \over d\C} \cdot \mbox{sign}(\C)d\C.
\label{intasym}
\end{equation}
Specifically we take the $\mbox{sign}(\C)$ to be zero if 
$\C=0$ to exclude that point, and use the following criterion 
to place bounds;
\begin{equation}
|\int \frac{d\sigma}{d\C}\cdot \mbox{sign}(\C) d\C| \leq 2\cdot \Delta 
\sigma_{exp}\;. 
\end{equation}
The right-hand side of this equation corresponds to two 
standard deviations for the experimentally measured cross-section. 
Given the definition of $A$ it is appropriate to use the same 
experimental uncertainty of the total cross-section:
\begin{equation}
\Delta \sigma_{exp}=\sigma_{SM}\cdot \sqrt{\delta^2_{syst}+
\delta^2_{stat}}\;, 
\end{equation}
where
\begin{equation}
\delta_{stat} = \frac{1}{\sqrt{ \sigma_{SM}\cdot \cal{L} \cdot 
\epsilon_{\mu}}}\;. 
\end{equation}
The best bounds come from using the relaxed cut of Eq.~(\ref{mincut}) 
and are given by
\begin{equation}
|\tilde k_Z + 0.7\cdot \tilde k_{\gamma} | \leq 0.2 
\label{intasymb}
\end{equation}
It is also possible to obtain bounds on $g_4^Z$ but they are much 
weaker.

\section{Conclusions}

We have studied the effect of $CP$ violating anomalous couplings 
on the process $e^+e^- \rightarrow W^+W^-$. Using $CP$ even observables 
we have found that a NLC with $\sqrt{s}=500$~GeV and $50$~pb$^{-1}$ can 
place the bounds $|\tilde\kappa_{\gamma,Z}| \leq 0.1$ 
and $|g_4^Z| \leq  0.09$. These bounds are comparable to those that 
can be placed with an upgraded Tevatron, and are of the same order 
as the bounds that can be placed on $CP$ conserving anomalous 
couplings. These bounds originate in the quadratic 
contributions of the couplings to the differential cross-section. 
By looking at the $\mu\nu$ decays of the $W$-bosons we were able to 
construct a $CP$ odd correlation that can directly bound the 
$CP$-violating terms (linear in the couplings) in the differential 
cross-section. We found that it will be possible to place the bounds 
$|\tilde\kappa_\gamma| \leq  0.3$ and  
$|\tilde\kappa_Z| \leq  0.2$. We conclude that
the sensitivity of a NLC to $CP$ violating anomalous couplings is 
similar to its sensitivity to $CP$ conserving anomalous couplings. 
From our dimensional analysis we also conclude that it is unlikely 
that $CP$ violating anomalous couplings will be seen by a NLC 
unless the smallness of $\Delta\rho$ is accidental.

\vspace{0.5in}
\noindent{\bf Acknowledgments}
The work of A.A.L and O.P.Y. was supported in part by 
RFBR under grant number 96-02-18216. The work of G.V. was supported in 
part by the DOE OJI program under contract number DEFG0292ER40730. We 
thank S. Dawson for comments on the manuscript.

\end{document}